**Full-count PET Recovery from Low-count Image Using a Dilated Convolutional Neural Network**

Karl Spuhler[1],[*],[†], Mario Serrano-Sosa[1],[*], Renee Cattell[1], Christine DeLorenzo[1,2], Chuan Huang[1,2,3]

[1] Department of Biomedical Engineering, Stony Brook University, Stony Brook, NY, USA
[2] Department of Psychiatry, Stony Brook University, Stony Brook, NY, USA
[3] Department of Radiology, Stony Brook University, Stony Brook, NY, USA

*Equal contribution

†Current Address: NYU Langone Medical Center, New York, NY, USA

Corresponding Author:

Chuan Huang, PhD
Department of Radiology
HSC-L4-120
Stony Brook Medicine
Stony Brook, NY 11794, USA
Email: Chuan.huang@stonybrook.edu

**Abstract**

Positron Emission Tomography (PET) is an essential technique in many clinical applications that allows for quantitative imaging at the molecular level. This study aims to develop a denoising method using novel dilated convolutional neural network to recover full-count images from low-count images. We adopted similar hierarchal structure from the conventional uNet and incorporated dilated kernels in each convolution to allow the network to observe larger, and perhaps, more robust, features within the image. Our dNet were trained alongside a uNet for comparison. Our 2.5D model used a training set (N=30) and testing set (N=5) that were obtained from an ongoing $_{18}$F-FDG study. Low-count PET data (10% count) were generated through Poisson thinning from the full listmode file. Both low-count PET and full-count PET were reconstructed with the OSEM algorithm. Objective imaging metrics including mean absolute percent error (MAPE), peak signal-to-noise ratio (PSNR) and structural similarity index metric (SSIM) were used to analyze the denoising methods. Both the uNet and our proposed dNet were successfully trained to synthesize full-count PET images from low-count PET images. Compared to low-count PET, both the uNet and dNet methods significantly improved MAPE, PSNR and SSIM. Our dNet also systematically outperformed uNet on all three metrics and across all testing subjects. This study proposed a novel approach of using dilated convolutions for recovering full-count PET images from low-count PET images. Our dNet significantly outperformed the well-established uNet and demonstrates great potential for denoising low-count PET images.

**1. Introduction**

Positron Emission Tomography (PET) is an integral part of contemporary cancer care. PET provides clinicians with a highly sensitive functional imaging tool to investigate a range of pathologies such as cancer, heart disease and brain disorders[1-4]. However, various physical degradation factors limit the number of detected photons, resulting in poor image resolution and signal to noise ratio (SNR). High SNR in PET images is advantageous for applications such as detection of small lesions and early diagnosis of neurological disease. In order to obtain images with high SNR for diagnostic or research use, the scanner must register a large number of radioactive decay events. As such, attaining high SNR comes at the cost of either relatively high radiation dose and/or long scan time[5]. Higher radiation doses can lead to greater risk of stochastic effects, such as greater lifetime risk of cancer[6]. Longer acquisition times require the subjects to lay still for an extended period of time, which increases the likelihood of subject motion that could degrade the image. Furthermore, although lowering dose or decreasing scan time would reduce patient burden, current reconstruction methods would result in images with unacceptable quality for diagnostic use[7].

The motivation to maximize SNR with low-count PET can offer great benefits, specifically to patient populations with cancer and researchers. Pediatric, adolescent and young adults with cancer may receive multiple scans for treatment monitoring or assessment of disease progression; they could greatly benefit from dose reduction through high SNR low-count PET [6,8]. Secondly, there are many instances in which research pursuits would desire to conduct multiple repeat scans on the same subject, such as for monitoring intermittent effects of psychiatric treatment along with treatment response. Given the limit for the amount of radiation exposure that research subjects can receive, researchers currently conduct primarily pre- and post-treatment PET scans on their subjects but often do not conduct multiple repeated scans. Developing a low-count PET denoising pipeline would increase the feasibility of projects with multiple repeated scans and allow researchers to observe neuropsychology at a deeper level.

Researchers have recently developed the first human total-body scanner, a hardware approach to improve SNR for low-count PET. This scanner, EXPLORER project [9], measures 2 meters in length and consists of 400,000 crystals--the equivalent of 8 conventional PET scanners--with the potential to obtain more counts than conventional PET scanners. With this enhancement in hardware, the total-body PET scanner has over 50 billion lines-of-response (LOR)[10]. This feature has allowed researchers to reconstruct high-quality PET images from 0.67 mCi (25 mBq or ~1/20 standard dose) with 6.9-fold increase in SNR [10] compared to conventional clinical PET. Although successful in improving SNR, hardware developments such as the total-body PET scanner are difficult to disseminate across clinical settings since it would cost clinics upwards of 20 million dollars.

Software advances are much more feasible to introduce into clinical workflow because it does not require costly new hardware installation. As such, three distinct software methods have been developed for improving the quality of PET images: iterative reconstruction algorithms, post-reconstruction image filtering and various machine learning methods. Iterative reconstruction methods have the desirable trait of operating with raw emission data[11]. In general, these algorithms treat low dose PET reconstruction as an optimization problem, where the goal is to estimate an image which would most likely lead to the raw data observed by the scanner. In addition to this, some manner of regularization is added to the image reconstruction objective to penalize noise properties. For example, iterative PET reconstruction algorithms have been augmented with a total variation regularizer, which seeks to enforce smoothness in the image space[12,13]. Despite the ideality of working in the raw data domain, iterative reconstruction algorithms suffer from increased computational time and are dependent upon many parameters that lack a principled method for selection.

Post-reconstruction methods that use image filtering or sparse methods to predict standard-dose PET from low-dose PET[14] have also succeeded in denoising PET images. Common image filtering techniques such as nonlocal means and block matching are well established in the field. Despite achieving higher visual quality, these methods tend to be critiqued for loss of sharp boundaries and are generally oversmoothed, which compromises textural information. Additionally, similar to iterative reconstruction approaches, they have reliance on a large number of parameters that are yet to be standardized.

Across all image denoising methods, proper assessment has been standardized for comparing the proposed method to ground truth. These assessments are typically done through the use of objective (i.e., quantitative assessment) and/or subjective (i.e., visual interpretation) measures, with the prior being more robust. Objective measures typically include mean absolute percent error (MAPE), peak signal to noise ratio (PSNR) and structural similarity index metric (SSIM). MAPE is a metric that distinguishes voxel-wise differences between ground truth and the proposed model; smaller differences depict an optimal denoising method. PSNR is a metric that measures the power of signal and power of corrupting noise; larger PSNR depicts an optimal denoising method. SSIM measures how well the proposed denoising method recovered structure and edge compared to the ground truth image; higher SSIM indicates a good denoising method. Subjective analysis uses visual interpretation of perceived image quality and the relative preservation of detail and edges to assess the performance of the denoising method.

Recently, machine learning methods for PET denoising have emerged and shown improvement in both objective and subjective assessment. Using a supervised dictionary-based method, Wang et.al successfully reduced dose by a factor of 4 with comparable image quality, as assessed by objective and subjective measures, to full-count PET[14,15]. Other methods using a modest convolutional neural network (CNN) architecture demonstrated increased PSNR compared to sparse-learning based procedures for recovering standard dose PET images[16]. Most strikingly, a publication proposed a residual uNet architecture which can reliably estimate standard dose PET data from low-dose scans with a dose reduction factor as high as 200[17]. Other machine learning frameworks have used multi-modal CNNs and end-to-end CNN reconstruction methods to estimate full-count PET from low-count PET. Chen et.al. used input of 2D PET slices with various 2D MRI contrasts such as T1, T2 and DWI into a uNet architecture to output full-dose PET images[18]. Although their results were successful in denoising the low-count image, development of a multi-modality method to denoise PET images is restrictive in accomplishing a generalizable PET denoising method that can be used across different scanners. Likewise, Chen's group used [$^{18}$F]-Florbetaben and a subjective scale for a binary (positive/negative) clinical evaluation task. Being a binary task, high resolution was not needed; therefore, the task was not sensitive to blurring, resolution, integrity of fine structure and edges. Häggström et.al. developed a deep encoder-decoder network for low-count PET reconstruction[19]. This work utilized simulated data and input sinograms into their network to output their simulated full-count PET image[20].

A majority of these studies have adopted the known and well-established uNet architecture. Typically, uNet architectures down-sample and up-sample feature maps as they are fed into the network, which degrades resolution and fine details. Notably, uNet processing introduces some degree of blurring from two primary sources. The first source is the mathematical nature of the convolution. Secondly, is the common practice of downsampling and subsequently re-upsampling feature maps as they pass through the network. Dilated kernels are a method to oppose this blurring effect. Recent development has brought about the use of these dilated kernels for dilated convolutions and have recently gained interest in the image segmentation community; wherein the dilated kernels enlarge the field-of-view to incorporate multi-scale context[21,22].

In this work, we propose to improve SNR of low-count PET brain images and predict full-count images by introducing a dilated convolutional neural network architecture (dNet) which is inspired by the uNet architecture. Our dNet uses dilated kernels that convolve with the feature maps, preserving resolution and simultaneously growing field of view to observe larger and more unique features. Residual learning will also be integrated into the architecture to capture the difference of low-count and full-count images and enhance convergence. The proposed model was implemented and evaluated on $^{18}$F-FDG PET images of the brain, with images reconstructed with 1/10th counts as input and original full-count images as output. The dilated convolution was originally introduced as a method to exponentially increase neuron receptive field size in a memory-efficient manner. We hypothesize that this paradigm would allow us to construct a novel multiscale dilated CNN approach which synthesizes sharper full-count PET estimates than the accepted uNet, as reflected by improved MAPE, PSNR and SSIM.

## 2. Methods
### 2.1 Method overview

Figure 1 displays the dilated convolution kernels used in dNet. The dilated convolution introduces a dilation factor to the standard convolution to define the amount of zero-padding placed between learnable elements of the filter. We developed two deep learning models for comparison of PET image denoising: a conventional uNet and our proposed dNet. Both models take low-count PET images as input and estimate full-count as the output. Comparison of these two models will be evaluated through objective imaging metrics: peak signal-to-noise ratio (PSNR), structural similarity index metric (SSIM) and mean absolute percent error (MAPE). Residual learning was also included into both networks, where the model assumed that the full count PET image can be expressed as the sum of the low-count PET data and a learned network representation.

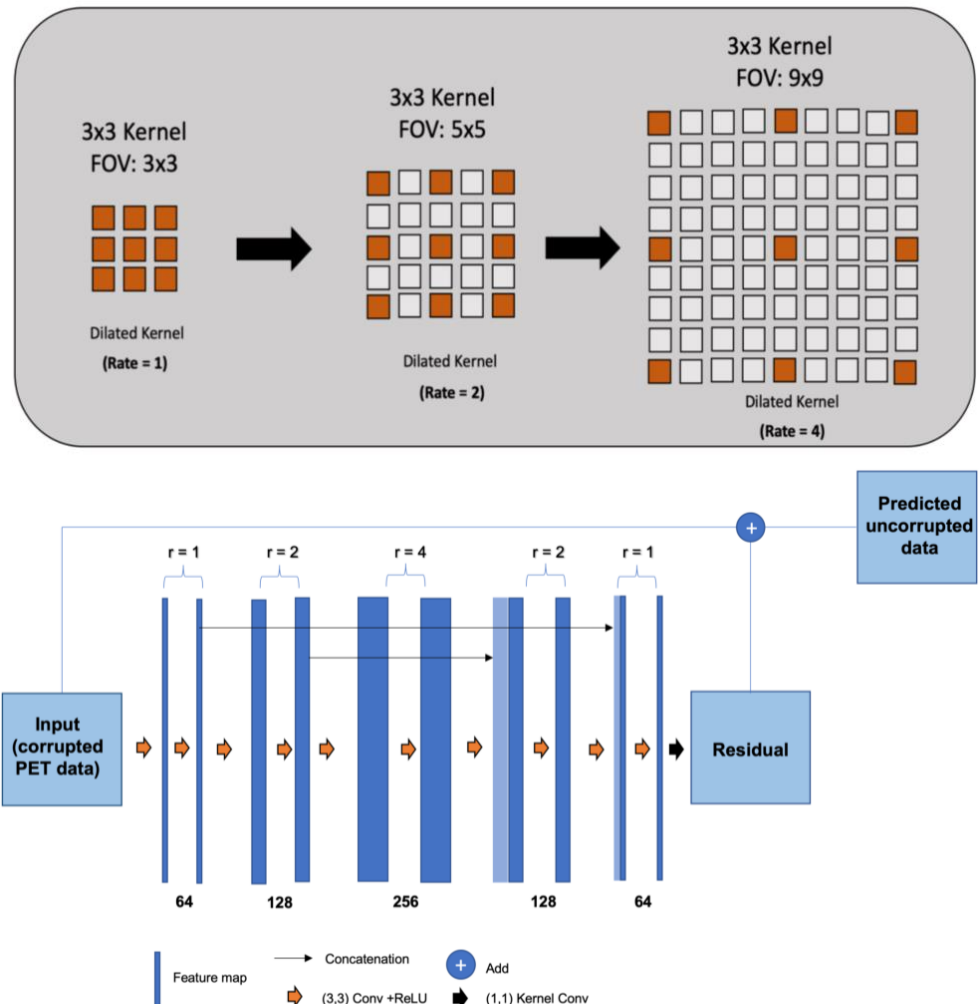

*Figure 1: Top: Illustration of dilated convolutions. Each operation can be represented using only 9 learnable weights but cover exponentially larger receptive fields given a linear increase in dilation rate. Bottom: Illustration of the dilated convolutional neural network, with residual learning, employed in this work.*

## 2.2 Dilated Convolution

Dilated convolutions were previously developed to improve segmentation tasks[21,22]. These dilated kernels are demonstrated in the top image of Figure 1. This dilation allows for enlarging field-of-view without increasing the number of parameters or the amount of computation and can potentially observe larger scaled features than typical static kernels. Dilated convolutions with rate $r$ introduces $r - 1$ zeros between consecutive filter values, effectively enlarging the kernel size of $k$ x $k$ filter to $k_e$ x $k_e$ wherein $k_e = k + (k-1)(r-1)$. This equates a compromise between accurate localization (small field-of-view) and context assimilation (large field-of-view)[21]. We utilized the exponentially expanding nature of dilated convolutions to conserve resolution in this PET denoising task. Our network employed these dilated convolutions in various rates from r = 1, 2, and 4.

2.3 Residual dNet training

Our proposed dNet architecture was inspired by uNet and keeps a similar multiscale hierarchal structure. 2D images are fed into dNet that is composed of repetitive densely connected convolutional blocks with static feature channel dimension but different $N$ feature channels between convolutional blocks; $N$ feature channels is double in the "encoder" path and halved in the "decoder" path. It consists of five convolutional blocks, in which each block has two 3 x 3 kernel-Convolutional layers followed by a rectified linear unit (ReLU) activation. Whereas uNet feature maps would contain maxpooling functions in its "encoder" path, or "transpose convolutions" in the decoder path, a dilation factor was incremented or decreased, respectively, in each block of convolutions as shown in the bottom image of Figure 1. This ultimately preserved resolution of the image across the entire path of the network. Lastly, the "decoder" path of dNet made use of skip connections as originally employed in uNet[23].

2.3 Residual uNet training

To determine if our dNet outperformed the already established uNet, we trained a conventional residual uNet with network architecture as shown in Figure 2. This uNet consisted of an encoding path (contracting, left side) and decoding path (expanding, right side). The encoding path conforms to the typical architecture of a CNN, consisting of the repeated application of two 3 x 3 convolution layers, each followed by a ReLU. Each block ended with a 2 x 2 maxpooling layer for downsampling followed by another 3 x 3 convolution layers + ReLU. In addition, at each downsampling step, the number of feature channels is doubled in the encoder path. The decoding path consists of a 3 x 3 transpose convolutional layer for upsampling, instead of the previously used maxpooling. In this decoding path, the feature channel is halved and skip connection with the corresponding linked feature map from the encoding path are utilized. The final step is a 1 x 1 convolution that maps the output residual.

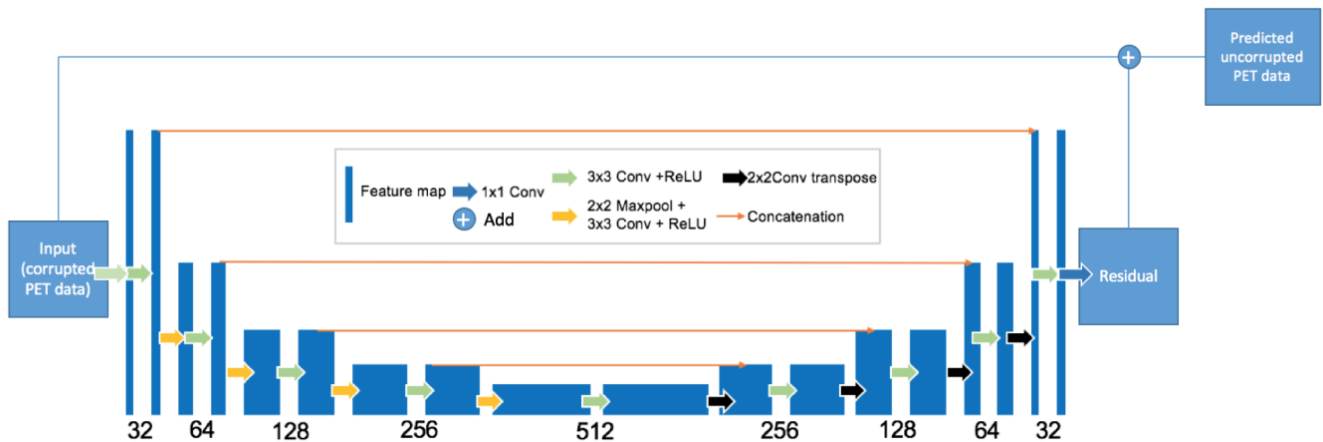

*Figure 2: Residual uNet architecture employed in this work*

## 2.4 Model Training

Both uNet and dNet models were trained using residual learning. All steps were virtually exactly the same in order to fairly compare the performance of the two models. Both models were trained to minimize the mean L1-error between their output and the ground truth. Both models were trained on 2D slices and, in order to afford the networks a degree of 3D information, the target slice and two adjacent inferior/superior slices were provided as independent input channels. Both models were trained using the Adam optimizer[24] and non-decaying learning rate of $1*10^{-5}$; network parameters were initialized using the Glorot method[25]. All convolutional kernel sizes were 3 x 3 and all convolutional layers other than the output layer of each network employed batch normalization. The two networks were trained for 200 epochs. Finally, the low count network input was multiplied by the dose reduction factor in order to accelerate network training and learn to scale the output by this factor. Both models were trained on a computer with an i9-7980XE 18-core processor, 128 GB memory, and two GTX 1080 Ti Graphic cards running Ubuntu 18.04, Python 2.7.15, TensorFlow 1.14.0.

## 2.5 Dataset

PET data for this project was extracted from an IRB approved, ongoing human study. A total of 35 PET studies were acquired and split into training (n=30) and testing (n=5). Each subject was administered between 148-185 MBq (4-5mCi) of $_{18}$F-FDG and asked to void their bladder immediately after injection. This ongoing study acquired listmode data collected using a dedicated head coil for 60 minutes immediately after the injection of $_{18}$F-FDG using a Siemens Biograph mMR PET/MRI scanner. Attenuation map were generated using an MRI-based algorithm, namely the "Boston Method"[26,27]. Scanner attenuation maps were also extracted for reconstruction. PET images were reconstructed using Siemens' E7tools with ordered subset expectation maximization (OSEM). Two sets of images were reconstructed using all 60 minutes of emission data and emission data acquired between 50 and 60 minutes after injection with the same parameters (OSEM: 6 iterations, 21 subsets) and attenuation map. The dNet and uNet models were trained twice using these two different sets (reconstructed images from 60-min emission and 10-min emission data).

## 2.6 PET Processing and Reconstruction

Data were prepared using Siemens e7tools package. Low-count PET data were generated through Poisson thinning. Specifically, low-count PET data with a dose reduction factor 90% (i.e. one-tenth of original counts) were generated. In each instance, two randomly sampled low-count listmode files were generated and reconstructed for each training set subject to increase the amount of training data. Figure 3 shows the sagittal, coronal and transverse views of low-count data and full-count data, where the low-count PET image appears grainy and extremely noisy.

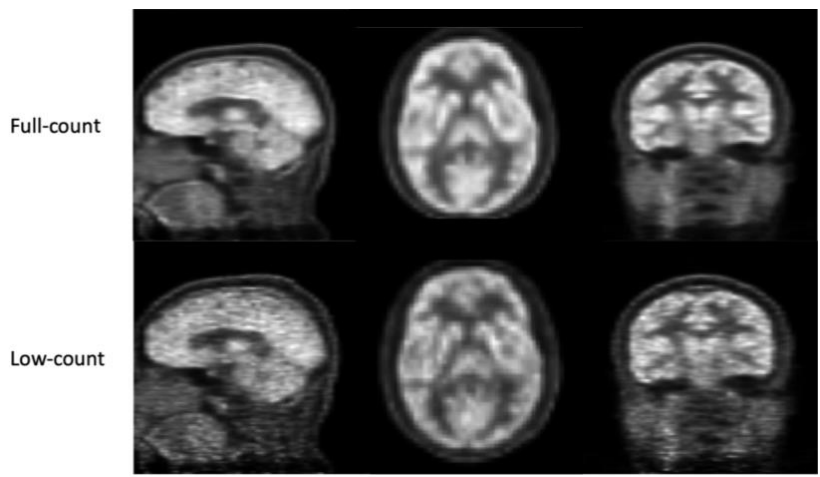

*Figure 3: Sagittal, coronal and transverse views of full-count data and low-count data (60-minute listmode) after being reconstructed using ordered subset expectation maximization (OSEM)*

## 2.7 Objective Image Quality Evaluation

The primary objective metric of the image quality in this study is the mean absolute percent error (MAPE) of the CNN-denoised data relative to the full count dataset. MAPE can be defined as:

$$MAPE(x, y) = \frac{1}{n} \sum_{i=1}^{n} \left| \frac{y_i - x_i}{y_i} \right|, \qquad (Eq.1)$$

where $y_i$ is the $i$th voxel in the ground truth image (y) and $x_i$ is the $i$th voxel in the denoised PET data.

Other quantitative image quality metrics widely accepted in the field were also studied, including peak signal-to-noise ratio (PSNR) and structural similarity index (SSIM) for the full-count reconstructed PET data and resultant denoised images. PSNR is an objective measure of image quality [28] defined as:

$$PSNR(x, y) = 20 * \log_{10} \left( \frac{MAX(y)}{\sqrt{MSE(x,y)}} \right), \qquad (Eq.2)$$

where y is the signal of the ground truth full-count PET data, x is the signal from the denoised PET data, MAX indicates maximum signal and MSE indicates the mean squared error between the two signal intensities.

SSIM is more complex and accounts for patch-wise image statistics and is defined as:

$$SSIM(x,y) = \frac{(2\mu_x\mu_y+c_1)(2\sigma_{xy}+c_2)}{(\mu_x^2+\mu_y^2+c_1)(\sigma_x^2+\sigma_y^2+c_2)}, \quad \text{(Eq.3)}$$

where y is the signal of the patch-wise ground truth full-count PET data, x is the signal of the patch-wise denoised PET data, $\sigma_x$ represents the variance of x, $\sigma_y$ represents the variance of y; $\sigma_{xy}$ represents the covariance of x and y; $\mu_x$ represents the mean of x, $\mu_y$ represents the mean of y and $c_1$ and $c_2$ are stabilizing terms.

## 3. Results

Both the uNet and our proposed dNet were successfully trained to synthesize full-count PET images from low-count PET images. Figure 4a shows i) the low-count image, ii) the image denoised by uNet, iii) the image denoised by our proposed dNet and iv) the image reconstructed from full count 60-min emission data. Upon subjective visual inspection, both uNet and dNet were able to improve the low-count image. As shown here, both CNN models yielded images similar to the images reconstructed with full-count data by removing the noise in the low-count images. The red arrows point to a region where the edge was difficult to differentiate in the low-count images but recovered in both CNN denoised images.

To better visualize the improvement afforded by dNet, line profiles of the yellow line shown in the figure is also provided in Figure 4b. As shown in the blue box, both CNN models recovered the activity that was lost in the low-count image. More notably, dNet yield line profiles more similar to the full-count curve compared to uNet.

Figure 4c, 4d show the same comparison for the 10-min emission data. As shown here, same observation can be made to images reconstructed from 60-min emission data (Figure 4a, 4b).

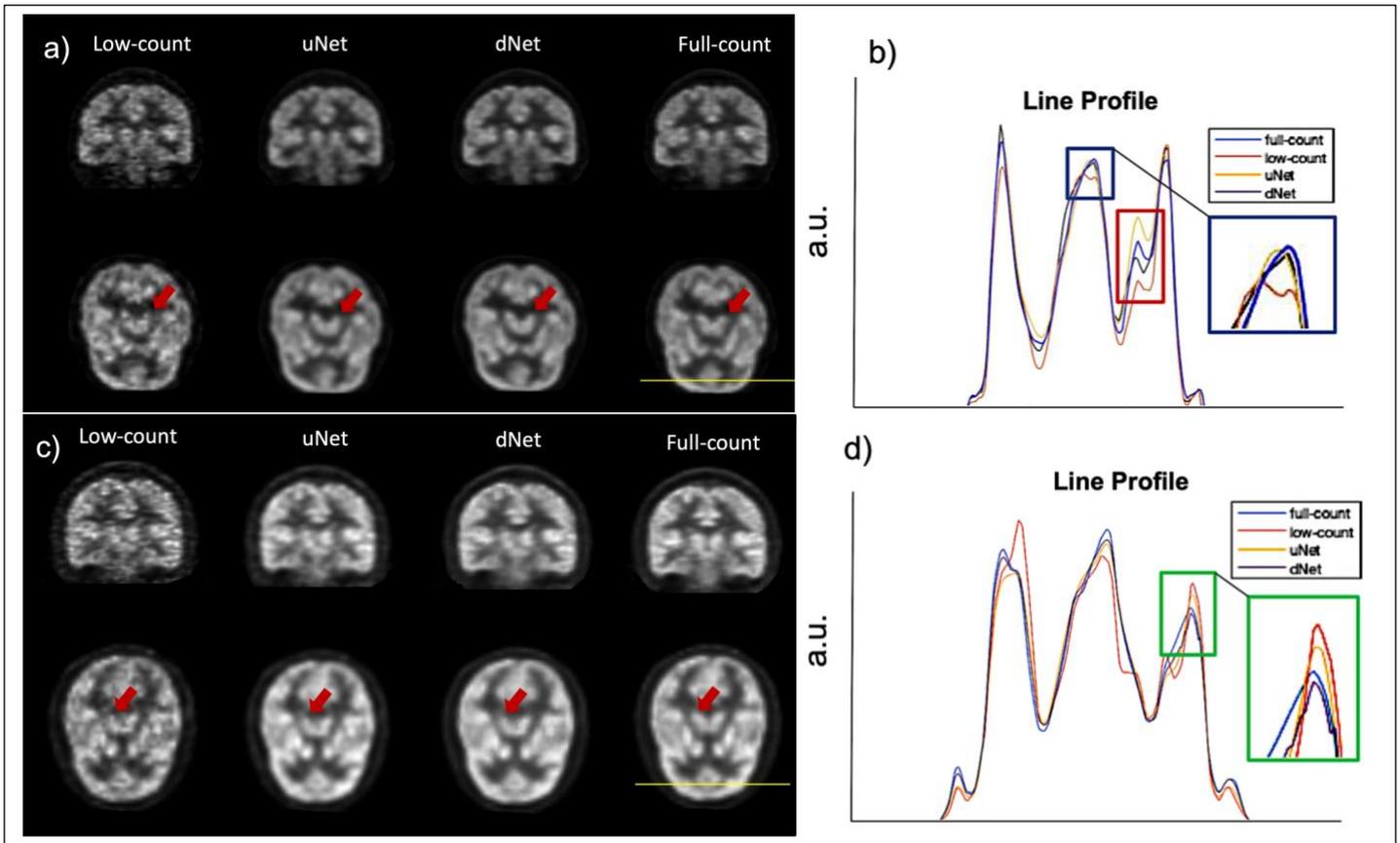

*Figure 4: Representative coronal and axial slices of low-count, uNet, dNet and full-count images. a) Images reconstructed from the 60-min emission, b) the line profile for the yellow line in a), c) images reconstructed from 10-min emission and d) the line profile for the yellow line in c). Red arrows point to a region where noise greatly reduced the fidelity of an edge in the low-count data, which both the CNN methods recovered in both 60-min and 10-min emission data sets. The blue box in b) displays a section in which both uNet and dNet recovered counts that were initially lost in the low-count data. The red box in b), and green box in d), show a section in which dNet performed better than uNet by having a smaller absolute difference compared to the full-count image.*

Mean and standard deviation of the objective imaging metrics for 60-min emission is shown below in Table 1. The first row represents objective measure of the low-count (1/10th count PET image) as compared to the ground truth (full-count PET image). Rows 2 and 3 show metrics calculated after the denoising of the testing set with the two different CNN models. An objective improvement in image quality is reflected by larger value in PSNR or SSIM and smaller value in MAPE. Our results demonstrate that uNet and dNet were both effective at denoising the low-count image. MAPE improved when comparing the low-count images to the uNet and dNet (6.57±0.38 vs. 5.15±0.22 and 4.96±0.23 [p<0.001, dNet compared to uNet], respectively). PSNR was also improved when comparing low-count images to uNet (35.90±0.74 dB vs 38.24±0.78 dB, p<0.01) and low-count to dNet (35.90±0.74 dB vs 38.67±0.78 dB, p<0.001). SSIM was also shown to be significantly improved from low-count images to uNet (0.89±0.02 vs 0.91±0.01, p<0.05) and dNet (0.89±0.02 vs 0.92±0.01, p<0.01). Using a paired samples t-test, our dNet model significantly outperformed uNet across all metrics (p<0.001).

Figure 5 shows the plotted image quality metrics for the reconstructed images from 60-min emission. The black lines connecting uNet to dNet data points in this figure show corresponding metrics for each testing set subject in the CNN models. Within each objective metric, dNet systematically outperformed uNet. Specifically, each subject had higher PSNR and SSIM along with lower MAPE in dNet compared to uNet.

Further analysis demonstrates the same behavior when using the 10-minute emission data (Table 2, Fig. 6).

| Model | Structural Similarity Index (SSIM) | Peak Signal-to-Noise Ratio (PSNR) | Mean Absolute Percent Error (MAPE) |
|---|---|---|---|
| Low-count | 0.89±0.02 | 35.90±0.74 dB | 6.57±0.38 |
| uNet | 0.91±0.01 | 38.24±0.78 dB | 5.15±0.22 |
| dNet | 0.92±0.01*** | 38.67±0.78 dB*** | 4.96±0.23*** |

*Table 1: Mean and standard deviation for testing set subjects in the reconstructed images from 60-min emission. All metrics were calculated relative to the full-count reconstruction. An objective improvement in image quality is reflected by larger value in PSNR or SSIM and smaller value in MAPE. (Statistically significant differences found between dNet and uNet using a paired t-test: \*\*\*p<0.001)*

| Model | Structural Similarity Index (SSIM) | Peak Signal-to-Noise Ratio (PSNR) | Mean Absolute Percent Error (MAPE) |
|---|---|---|---|
| Low-count | 0.84±0.01 | 24.96±0.38 dB | 10.6±0.88 |
| uNet | 0.87±0.01 | 26.92±0.73 dB | 8.61±0.63 |
| dNet | 0.89±0.01*** | 27.70±0.66 dB** | 7.93±0.59** |

*Table 2: Mean and standard deviation for testing set subjects in the reconstructed images from 10-min emission. All metrics were calculated relative to the full-count reconstruction. An objective improvement in image quality is reflected by larger value in PSNR or SSIM and smaller value in MAPE. (Statistically significant differences found between dNet and uNet using a paired t-test: \*\*p<0.01; \*\*\*p<0.001)*

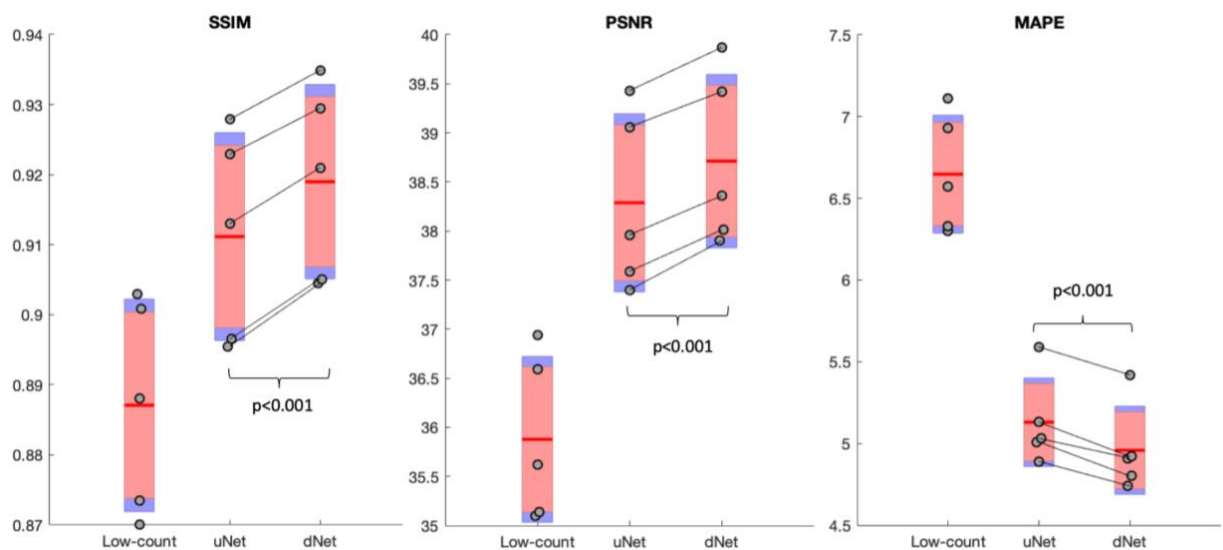

*Figure 5: Image quality metrics for images reconstructed from 60-min emission. In the task of PET denoising, for each subject, dNet significantly outperformed uNet across all metrics (p<0.001)(Summary data provided in Table 1).The box in red indicates the 95% confidence interval (CI) and blue represents one standard deviation, the red lines represent the mean, circles indicate the measured data for each subject. The lines connecting uNet and dNet show corresponding subjects. An objective improvement in image quality is reflected by larger value in PSNR or SSIM and smaller value in MAPE.*

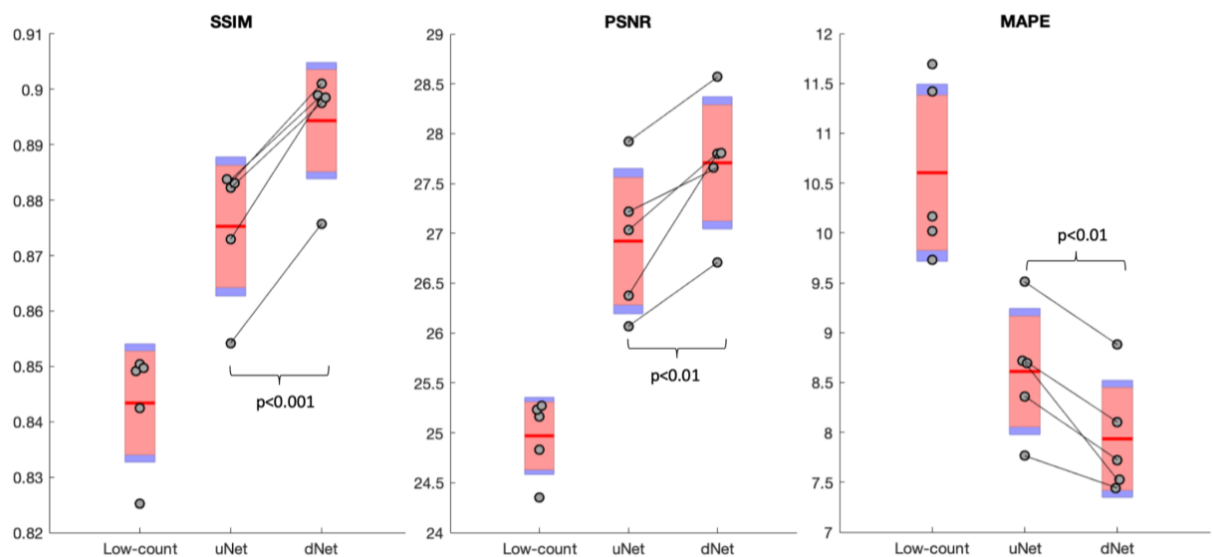

*Figure 6: Image quality metrics for images reconstructed from 10-min emission. For each subject, dNet significantly outperformed uNet across all metrics (p<0.01)(Summary data provided in Table 2). The box in red indicates the 95% confidence interval (CI) and blue represents one standard deviation, the red lines represent the mean, circles indicate the measured data for each subject. The lines connecting uNet and dNet show corresponding subjects. An objective improvement in image quality is reflected by larger value in PSNR or SSIM and smaller value in MAPE.*

## 4. Discussion

In this work, we developed a dilated convolutional neural network, termed as dNet, to improve image quality in low-count PET images and compared it to well-established conventional uNet architecture. Both CNN models significantly improved image quality across several objective metrics and recovered structure and edges, as compared to degraded low-count images. The uNet and the proposed dNet were trained twice: 60-min emission data and 10-min emission data. For both iterations, dNet significantly outperformed uNet across all objective image quality metrics. The 10-minute emission data was acquired between 50-60 minutes after injection of [18]F-FDG; this is directly comparable to clinical FDG imaging. Therefore, dNet has shown to be an effective means for denoising low-count PET images that could enable applications such as low-dose PET. In recent studies, the standard uNet has been used to denoise low-count PET image and achieved promising results [17]. The present study replicated the results that uNet is a suitable tool for low-count PET denoising. More importantly, this study demonstrates that our novel application of a residual dNet architecture for PET denoising can categorically outperform the objective results provided by conventional uNet.

Dilated convolutions have been utilized in various segmentation tasks[21,22], but this is the first study to use dilated convolutions for low-count PET denoising. The adopted method inspired jointly by uNet and the notion of dilated convolutions has outperformed its predecessor in the task of providing sharper, full-count PET estimation from low-count PET images. Upon subjective visual inspection, denoised images depict boundaries and edges which are recovered from the low-count PET data. This demonstrates that dNet is a feasible way to denoise low-count PET data. Objectively, our results show that dNet was able to significantly improve image quality in low-count PET images with higher improvement in image quality on an intrapatient basis as compared to the already established and well-recognized uNet. This improvement is hypothesized to be the case because dilated convolutions are able to overcome the tradeoff between receptive field size and necessary downsampling of feature maps, which is a limitation of the state of the art uNet. The statistically significant, monotonic increases in image quality observed within each subject suggests that this technique should be further examined and improved as a method for edge preserving generative models.

Image blurring is a major limitation of PET denoising algorithms and methods to overcome this have received considerable attention[29,30]. It is encouraging that in PET analyses both CNN methods visually improved image quality, and that dNet significantly outperformed the objective metrics MAPE, PSNR and SSIM compared to the uNet method. With denoising methods, it is possible to reduce patient burden by reducing radiation exposure and shortening acquisition time. This would help various populations, specifically the pediatric oncology population. Retrospective studies have shown that pediatric patients will have on average 3.2 PET/CT scans throughout the diagnosis and treatment of their malignancy. Specifically, this patient population can average a cumulative effective dose of 1450 mrem throughout their treatment; this is four times greater than the annual radiation exposure to U.S. residents[31,32]. Likewise, very young children often require sedation for PET studies due to an inability to lie still for the duration of the acquisition period; this sedation is a concern for children, with a statistically significant increase in the chance of developing mental disorders, developmental delay and attention-deficient/hyperactivity disorder (ADHD)[33,34].

This project suggests several promising avenues for continued work. Existing work employing the lowest ground truth reference data [14] developed a largely hand-crafted approach which used incremental refinements of PET patches based on feature representations shared between PET, fractional anisotropy, mean diffusivity and T1w MRI data. Although the method achieved encouraging results for the relatively low count data employed, it utilized diffusion MRI data, which are known to differ significantly between scanners, even when the same scanner model and parameters are employed[35]; furthermore, it is also significantly more difficult to acquire images from multiple modalities. Our methods add to a growing trend in contrast to previously existing standards which require additional data or selection features to improve low-count PET image quality[13].

This study was subject to some limitations. In particular, an expanded dataset would allow for better characterization of uNet and dNet as denoising tools. Most notably, acquiring sufficient data for a training-validation-testing paradigm would allow for unbiased future developments in network structure and hyperparameter optimization.

## Conclusion
Our novel approach of using a dilated convolutional neural network architecture for low-count PET denoising is an accurate way to improve SNR and recover full-count PET images. Concurrently, dNet significantly outperformed the conventional residual uNet trained using the same dataset. Further denoising machine learning methods could utilize the spatially conserving aspect of dNet to improve their results.

## Acknowledgement
This work is in part supported by NARSAD Young Investigator Award (26144, Huang), R01MH104512 (DeLorenzo). The authors would like to thank Siemens Healthcare for providing E7Tools.
## Disclosure
All authors declare that they have no relevant conflicts.